\documentclass[useAMS,usenatbib]{mn2e}
\usepackage{amssymb,amsmath,graphicx}
\usepackage{longtable}[1]
\usepackage{graphics}
\usepackage{threeparttable}
\usepackage{lscape}
\usepackage{color}
\title[The birth of a cluster complex]{Revealing a Ring-like Cluster Complex in a Tidal Tail\\ of the Starburst Galaxy NGC 2146\thanks{Based on observations made with the NASA/ESA Hubble Space Telescope, obtained at the Space Telescope Science Institute, which is operated by the Association of Universities for Research in Astronomy, Inc., under NASA contract NAS 5-26555. These observations are associated with program SNAP 12229}}

\author[Angela Adamo et al.]{A.~Adamo$^{1}$\thanks{E-mail:
adamo@mpia.de}, L.~J.~Smith$^{2}$, J.~S.~Gallagher$^{3}$, N.~Bastian$^{4}$, J.~Ryon$^{3}$,  \newauthor M.~S.~Westmoquette$^{5}$,  I.~S.~Konstantopoulos$^{6}$,  E.~Zackrisson$^{7}$, S.~S.~Larsen$^{8}$,  \newauthor E.~Silva-Villa$^{9}$, J.~C.~Charlton$^{6}$, D.~R.~Weisz$^{10}$\\
$^{1}$Max-Planck-Institut for  Astronomy, K\"onigstuhl 17, D-69117 Heidelberg, Germany\\
$^{2}$Space Telescope Science Institute and European Space Agency, 3700 San Martin Drive, Baltimore, MD 21218, USA\\
$^{3}$Department of Astronomy, University of Wisconsin-Madison, 5534 Sterling, 475 North Charter Street, Madison WI 53706, USA\\
$^{4}$Excellence Cluster Universe, Boltzmann-Strasse 2, 85748 Garching bei M\"unchen, Germany\\
$^{5}$ESO, Karl-Schwarzschild-Strasse 2, 85748 Garching bei M\"unchen, Germany\\
$^{6}$Department of Astronomy \& Astrophysics, The Pennsylvania State University, University Park, PA 16802, USA\\
$^{7}$Department of Astronomy, Stockholm University, Oscar Klein Center, AlbaNova, Stockholm SE-106 91, Sweden\\
$^{8}$Department of Astrophysics/IMAPP, Radboud University Nijmegen, P.O. Box 9010, 6500 GL Nijmegen, The Netherlands\\
$^{9}$D\'epartement de Physique, de G\'enie Physique et d'Optique, and Centre de Recherche en Astrophysique du Qu\'ebec (CRAQ),\\
 Universit\'e Laval, Qu\'ebec, Canada\\
$^{10}$Department of Astronomy, Box 351580, University of Washington, Seattle, WA 98195, USA\\}

\newcommand{\araa}{ARA\&A}
\newcommand{\apj}{ApJ}
\newcommand{\aj}{AJ}
\newcommand{\mnras}{MNRAS}
\newcommand{\aap}{A\&A}
\newcommand{\pasp}{PASP}

\newcommand{\apjs}{ApJS}
\newcommand{\msun}{M_{\odot}}

\begin{document}

\date{Accepted 2012 May 23.  Received 2012 May 23; in original form 2012 March 26}

\pagerange{\pageref{firstpage}--\pageref{lastpage}} \pubyear{2012}

\maketitle

\label{firstpage}

\begin{abstract}

We report the discovery of a ring-like cluster complex in the starburst galaxy NGC\,2146. The Ruby Ring, so named due to its appearance, shows a clear ring-like distribution of star clusters around a central object. It is located in one of the tidal streams which surround the galaxy. NGC\,2146 is part of the Snapshot Hubble $U$-band Cluster Survey ({\it SHUCS}). The WFC3/F336W data has added critical information to the available archival Hubble Space Telescope imaging set of NGC\,2146, allowing us to determine ages, masses, and extinctions of the clusters in the Ruby Ring. These properties have then been used to investigate the formation of this extraordinary system. We find evidence of a spatial and temporal correlation between the central cluster  and the clusters in the ring. The latter are about 4 Myr younger than the central cluster, which has an age of 7 Myr. This result is supported by the H$\alpha$ emission which is strongly coincident with the ring, and weaker at the position of the central cluster. From the derived total H$\alpha$ luminosity of the system we constrain the star formation rate density to be quite high ($\Sigma_{\textnormal{SFR}}=0.47 \msun$/yr/kpc$^{2}$). The Ruby Ring is the product of an intense and localised burst of star formation, similar to the extended cluster complexes observed in M\,51 and the Antennae, but more impressive because is quite isolated. The central cluster contains only 5 \% of the total stellar mass in the clusters that are determined within the complex. The ring-like morphology, the age spread, and the mass ratio support a triggering formation scenario for this complex. We discuss the formation of the Ruby Ring in a "collect \& collapse" framework. The predictions made by this model agree quite well with the estimated bubble radius and expansion velocity produced by the feedback from the central cluster, making the Ruby Ring an interesting case of triggered star formation.

\end{abstract}

\begin{keywords}
galaxies: individual: NGC\,2146 -- galaxies: starburst -- galaxies: star clusters: general -- galaxies: star formation
\end{keywords}

\section{Introduction}


Star formation appears to be a hierarchical process \citep[see][for a short review]{2011EAS....51...31E}, organised as a fractal filamentary network, self-similar from small (dense gas cores inside a giant molecular cloud, GMC) to galactic scales (GMC complexes). Much observational and theoretical evidence has been put forward to support this scenario. Pioneering studies of the 2 and 3-dimensional structures of molecular clouds suggested a fractal medium \citep{1998A&A...336..697S}. The superb observations at sub-mm and far-IR, provided by {\it Herschel}, recently confirmed the filamentary and hierarchical structures of the regions which will most likely host star formation in the near future \citep[e.g. the GMC complex Vela C, ][]{2011A&A...533A..94H}. The same behaviour is observed in nearby star-forming galaxies \citep{2008MNRAS.391L..93G, 2009A&A...494...81S, 2011MNRAS.412.1539B}, where the very young stellar population is clearly clustered, while a gradual dissolution of the network is observed as the stars age.

In this hierarchical network, star clusters sit at the bottom as the densest objects of a continuous distribution which includes distributed stellar agglomerates, T Tauri like associations, and OB associations \citep{2010MNRAS.409L..54B}. On large scales, since clusters form in the dense cores of the GMCs, they can be part of star-forming complexes. These extended (few hundreds of pc) structures are often observed in the Milky Way \citep{2008hsf1.book..459B, 2010ApJ...719.1104R} and in local galaxies \citep[e.g.][]{2000ApJ...535..748E, 2001ApJ...561..727Z, 2005A&A...443...79B}. In particular, shells and arc-like star forming complexes seem to show a spatial and temporal sequentiality, with older systems sitting toward the centre \citep{1985ApJ...297..599D, 1998ApJ...507..241P, 1998MNRAS.299..643E, 2006A&A...446..171Z}. The arc-like morphology can be explained by the "collect and collapse" triggering mechanism \citep[][]{1977ApJ...214..725E, 1998ASPC..148..150E}. In this scenario, the expansion of an ionised shell, powered by stellar feedback, accumulates neutral gas, left over from the formation of the first generation of stars. The gravitational instability, induced on the swept up material at the interface with the ionised front, triggers a subsequent star formation episode.

\begin{figure*}
\resizebox{\hsize}{!}{\rotatebox{0}{\includegraphics{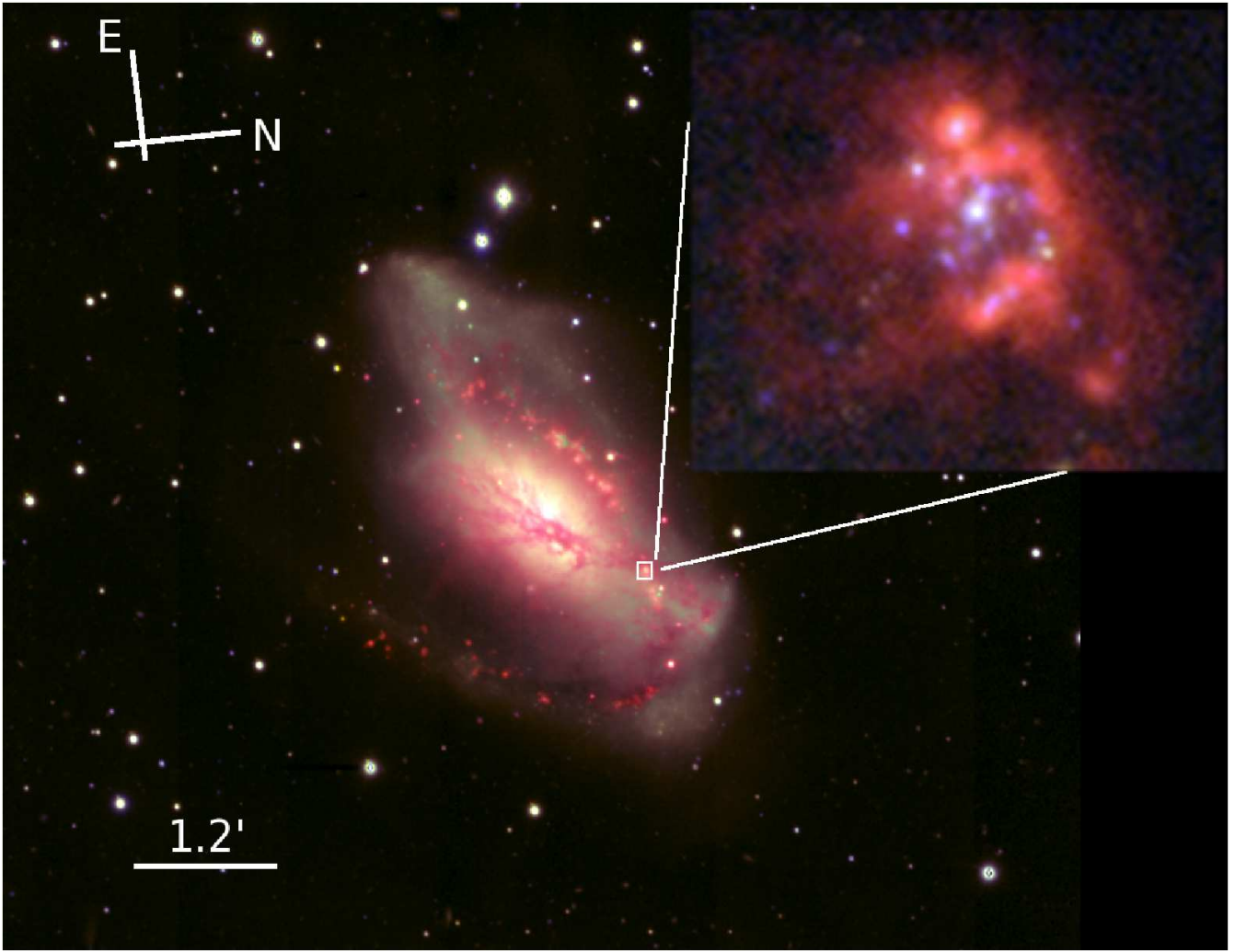}}}\\
\caption{A color composite of NGC\,2146, using WIYN $B$, $R$, and continuum subtracted H$\alpha$ imaging data in the blue, green, and red channels, respectively. An inset in the upper right corner shows a zoom-in image of the Ruby Ring region, made with {\it HST} data. In the latter, the F658N filter, transmitting H$\alpha$, is in red; the F336W filter, sensitive to the young stellar populations is shown in blue; the F606W and F814W are shown in green and yellow, respectively. The size of the inset is 2.5" by 2.4" ($\sim130\times125$~pc$^2$ at the assumed distance of 11~Mpc).}
\label{3col}
\end{figure*}

In this work, we report the discovery of a ring-like cluster complex (see Figure~\ref{3col}), in the tidal stream of the nearby starburst galaxy NGC\,2146  \citep[$\sim11$~Mpc,][]{rc3}. The appearance of this feature has led us to name it the "Ruby Ring". The Ruby Ring is located toward the extremity of one of the tidal streams. The latter has most likely been ejected during a merger event \citep{2001A&A...365..360T}. The properties of the stream and the analysis of the cluster population in the whole galaxy will be presented in a subsequent work (Adamo et al. in prep). Here, we focus on the properties of the Ruby Ring and its immediate environment, with the aim of determining whether its formation can be explained by the triggering scenarios applied to much smaller, and less massive structures seen in the Galaxy and Large Magellanic Cloud \citep{2012MNRAS.tmp.2286T, 1998MNRAS.299..643E}. We derive ages, masses and extinctions for the cluster population of the Ruby Ring, and consider whether the central cluster could have triggered the formation of the surrounding ring of clusters.

\section[]{Reduction and photometry}
\label{data-sample} 

The dataset used here forms part of the "Snapshot Hubble U-band Cluster Survey (SHUCS)", which is aimed at measuring the properties of star clusters in nearby star-forming galaxies using archival {\it HST} BVIH$\alpha$ and new U-band WFC3 data. The first paper (Konstantopoulos et al., in prep) presents the sample, and the reduction, cluster selection and photometric analysis steps. In a second paper (Adamo et al. in prep), we focus on the cluster population of NGC 2146, and here we use the subset of the {\it HST} data that contains the Ruby Ring complex. These datasets are: WFPC2 F606W (ID 8597, PI Regan), ACS WFC F814W, F658N (ID 9788, PI Ho), and WFC3/UVIS F336W (hereafter also referred to as $U$ band -- ID 12229, PI Smith). Unfortunately, the Ruby Ring fell out of the WFPC2 F450W field of view, limiting our sample to 4 passbands, i.e. UVIH$\alpha$.

The final science frames, used in this analysis, have been processed with the {\tt MULTIDRIZZLE} task \citep{2002hstc.conf..337K, 2002PASP..114..144F} in {\tt PyRAF/STSDAS}\footnote{STSDAS and PyRAF are products of the Space
Telescope Science Institute, which is operated by AURA for NASA}. Source detection has been done with the {\tt DAOFIND}  task on the $U$-band frame, which has the best resolution (0.04"/px plate scale, corresponding to $\sim 2$ pc at the assumed distance) and detection limit ($\sim$26.0 mag). The Ruby Ring catalogue contains 25 sources, all detected inside a radius of 1.25" ($\sim$ 80 pc) from the central cluster (hereafter indicated as 180). This catalogue has been used to do photometry with an aperture of 0.12" and the sky annulus located at 0.2" in all the available filters. Since not all the science frames have stellar-like objects which we could use to extrapolate an empirical aperture correction, we used {\tt TinyTim}\footnote{http://www.stsci.edu/software/tinytim/} simulated PSFs to estimate the fraction of missing flux in all the passbands. In this way, since the distance of the galaxy is such that stellar clusters are partially resolved (slightly more extended than stars), we may have underestimated the aperture correction factor and, thus, the estimated cluster mass. In Adamo et al. (in prep), we will present a full detailed analysis related to this aspect. However, we stress that the overall analysis in Section 3 and 4 appears to be consistent with the derived clusters properties, suggesting that we don't introduce significant systematics. 

The final photometry and derived errors of the cluster candidates are listed in Table~\ref{table1}. We notice that the observed absolute luminosity of the sources is comparable to the one of supergiant stars.  However, the observed H$\alpha$ emission (Section~\ref{ciaociao}) within the ring is too high to be produced by only 10--20 single supergiants (the detected sources visible in the complex). Although, we cannot exclude that a fraction of the detected sources are single stars, our analysis suggests that their contamination is a minor effect. In Section~\ref{ciaociao} we will see that the derived H$\alpha$ luminosity, L(H$\alpha$), is comparable to the 30 Doradus region, requiring a powering source of $\sim 10^5 \msun$. 

\begin{table*}
  \caption{Detected cluster candidates in the Ruby Ring. The first 5 columns give the ids as used in the text, the absolute magnitude in F606W applying a distance modulus of 30.38 mag, and  the final apparent magnitude of the objects (including aperture correction and correction for foreground extinction) and their associated error, for each filter, in the Vega system. All the photometry has not been corrected by the intrinsic extinction derived with the SED fitting (last column of the table). The FLT column enumerates how many data-points have been included in the SED fit (detections with photometric errors larger than 0.3 mag have been excluded). The $\chi^2$ is the best (e.g. smallest) recovered $\chi^2$ output. The following columns show the age, mass, and E(B-V) corresponding to the best $\chi^2$ and, where possible, the associated fit uncertainties  (see text). Empty cells indicate that the fit has not been performed for these sources, because there
were too few data-points for an accurate SED fit.}
\centering
  \begin{tabular}{|c|c|c|c|c|c|c|c|c|c|c|}
  \hline
  ID & M$_V$&F336W &  F606W &  F658N & F814W & FLT &$\chi^2$ &age & mass & E(B-V) \\
   &  &WFC3 & WFPC2 &  ACS & ACS & & & & & \\
   \hline
 &(mag)&(mag)&(mag)&(mag)&(mag)&&& (Myr) & $10^3$ ($\msun$)& (mag)\\
 \hline
 \hline

     162&$-7.08$&23.45$\pm$0.09&23.30$\pm$0.17&20.18$\pm$0.09&23.02$\pm$0.12&       4&0.16&1.0$^{+0.0}_{-0.0}$ &5.05$^{+1.26}_{-1.10}$ &0.63
$^{+0.06}_{-0.06}$\\

     163&$-7.64$&23.96$\pm$0.14&22.74$\pm$0.12&23.02$\pm$0.82&22.20$\pm$0.07&       3&0.08&2.0 &21.28 &0.96\\

     164&$-7.40$&23.33$\pm$0.12&22.98$\pm$0.15&21.19$\pm$0.26&23.16$\pm$0.14&       4&4.52&5.0$^{+0.0}_{-0.0}$ &3.37$^{+1.15}_{-0.87}$ &0.48$^{+0.08}_{-0.07}$\\

     165&$-$7.64&23.82$\pm$0.09&22.74$\pm$0.12&22.26$\pm$0.36&22.18$\pm$0.07&       3&0.01&3.0 &16.26 &0.93\\

     166&$-$7.45&23.06$\pm$0.08&22.93$\pm$0.13& --&23.28$\pm$0.12&       3&0.05&2.0 &2.25 &0.38\\

     168&$-$7.89&22.45$\pm$0.07&22.49$\pm$0.11&20.39$\pm$0.11&22.63$\pm$0.09&       4&5.38&2.0$^{+0.0}_{-0.0}$ &4.23$^{+0.73}_{-0.67}$ &0.49$^{+0.04}_{-0.04}$\\

     169&$-$7.34&24.17$\pm$0.15&23.03$\pm$0.13&20.46$\pm$0.09&24.75$\pm$0.45&       3&0.05&2.0 &20.42 &1.03\\

     170&$-$6.50&25.22$\pm$0.40&23.88$\pm$0.46&21.08$\pm$0.07&25.14$\pm$0.65&       1&--&--&-- &--\\

     171&$-$5.49&24.55$\pm$0.26&24.89$\pm$0.73&21.58$\pm$0.12&23.41$\pm$0.14&       3&0.22&2.0 &3.99 &0.82\\

     172&$-$6.66&23.12$\pm$0.08&23.72$\pm$0.28&21.12$\pm$0.14&24.14$\pm$0.25&       4&0.60&3.0$^{+0.0}_{-1.0}$ &0.52$^{+0.20}_{-0.18}$ &0.23$^{+0.07}_{-0.12}$\\

     174&$-$7.24&25.44$\pm$0.82&23.14$\pm$0.11& --& --&       1&--&--&-- &--\\
     
     175&$-$8.94&23.20$\pm$0.12&21.44$\pm$0.06& --&22.83$\pm$0.10&       3&98.28& $< 1$ &7.71 &0.54\\

     177&$-$8.43&24.14$\pm$0.29&21.95$\pm$0.08&23.50$\pm$0.54&23.54$\pm$0.19&       3&68.74&$< 1$ &5.04 & 0.57\\

     178&$-$7.65&23.52$\pm$0.17&22.73$\pm$0.21& --&23.87$\pm$0.21&       3&4.89&$< 1$&1.65 & 0.35\\

     179&$-$7.23&23.00$\pm$0.08&23.15$\pm$0.19&22.76$\pm$0.42&23.11$\pm$0.10&       3&0.01&1.0 &3.40 &0.49\\

     180&$-$8.97&21.05$\pm$0.05&21.41$\pm$0.06&21.10$\pm$0.11&21.22$\pm$0.06&       4&2.33&7.0$^{+86.0}_{-0.0}$ &4.52$^{+30.08}_{-0.50}$ &0.10
$^{+0.03}_{-0.10}$\\

     181&$-$8.97&21.85$\pm$0.07&21.41$\pm$0.06&22.32$\pm$0.56&22.36$\pm$0.10&       3&18.97&$< 1$ &6.84 & 0.33\\

     182&$-$7.00&23.69$\pm$0.10&23.38$\pm$0.32&20.34$\pm$0.08&23.91$\pm$0.16&       3&0.01&$< 1$ &1.44 &0.36\\

     183&$-$8.43&23.94$\pm$0.20&21.95$\pm$0.08&22.63$\pm$0.44& --&       2&--&--&-- &--\\

     187&$-$6.20&23.97$\pm$0.20&24.18$\pm$1.00&20.83$\pm$0.09&23.36$\pm$0.25&       3&0.10&4.0 &3.73 &0.62\\

     188&$-$7.05&22.66$\pm$0.06&23.33$\pm$0.15&22.42$\pm$0.31&23.33$\pm$0.08&       3&0.02&4.0 &1.54 &0.30\\

     190&$-$8.75&22.22$\pm$0.06&21.63$\pm$0.06&18.80$\pm$0.05&21.79$\pm$0.06&       4&13.63&3.0$^{+0.0}_{-0.0}$ &11.49$^{+1.27}_{-1.57}$ &0.57
$^{+0.03}_{-0.04}$\\

     192&$-$7.74&22.75$\pm$0.07&22.64$\pm$0.14&21.59$\pm$0.30&22.16$\pm$0.06&       4&0.16&5.0$^{+1.0}_{-0.0}$ &9.14$^{+1.36}_{-6.50}$ &0.59
$^{+0.04}_{-0.23}$\\

    3156&$-$7.16&23.11$\pm$0.09&23.22$\pm$0.30&20.05$\pm$0.09&23.03$\pm$0.12&       4&0.34&1.0$^{+0.0}_{-0.0}$ &4.34$^{+1.11}_{-0.97}$ &0.55
$^{+0.06}_{-0.06}$\\

    3159&$-$7.13&23.75$\pm$0.19&23.25$\pm$0.43&21.40$\pm$0.15&24.48$\pm$0.60&       2&--&-- &-- &--\\
  \hline
\end{tabular}
\label{table1}
\end{table*}

\section{Tracing the physical properties of the cluster complex}

\subsection{Cluster age and extinction across the ring structure}
\label{age_sect}
In order to obtain an estimate of the cluster age, mass, and extinction, the photometric catalogue has been used as input for the $\chi^2$ fitting algorithm discussed in \citet{2010MNRAS.407..870A}. We compared the observed cluster spectral energy distribution (SED) to Yggdrasil models \citep{2011ApJ...740...13Z}, scaled by increasing values of extinction \citep{2000ApJ...533..682C}. These models are based on Starburst99 stellar population spectra \citep{1999ApJS..123....3L, 2005ApJ...621..695V} compiled with a Kroupa initial mass function \citep[IMF;][]{2001MNRAS.322..231K}, a range of metallicities (1/2, 1, and 2 times the solar metallicity), a gas covering factor of 0.5 and hydrogen density typical of local Galactic H{\sc ii}-regions (100 cm$^{-3}$). Only sources which have been detected in at least 3 filters, with errors less than 0.3 mag in each band, have been fitted. The narrow band H$\alpha$ filter, F658N, has been included in the fit. The cluster properties, where available, are listed in Table~\ref{table1}, together with the photometric catalogue. We have also investigated the 68 \% confidence limits on the determination of the cluster parameters (associated errors in Table~\ref{table1}), if the fitted data points (N$=$4) exceeded the number of fitted parameters (p$=$3), according to the prescription given by \citet{1976ApJ...208..177L}. In the latter case, the SED fitting analysis for each source with detection in all the available filters is shown in the Appendix.  
The analysis we present below is based on clusters with $\chi^2$ smaller than 20.0 and F606W-F814W color larger than $-0.9$ mag (to remove outliers with unphysical properties). For completeness, in Table~\ref{table1} we report the outputs of the whole photometric and SED fitting analysis.

\begin{figure*}
\resizebox{0.49\hsize}{!}{\rotatebox{0}{\includegraphics{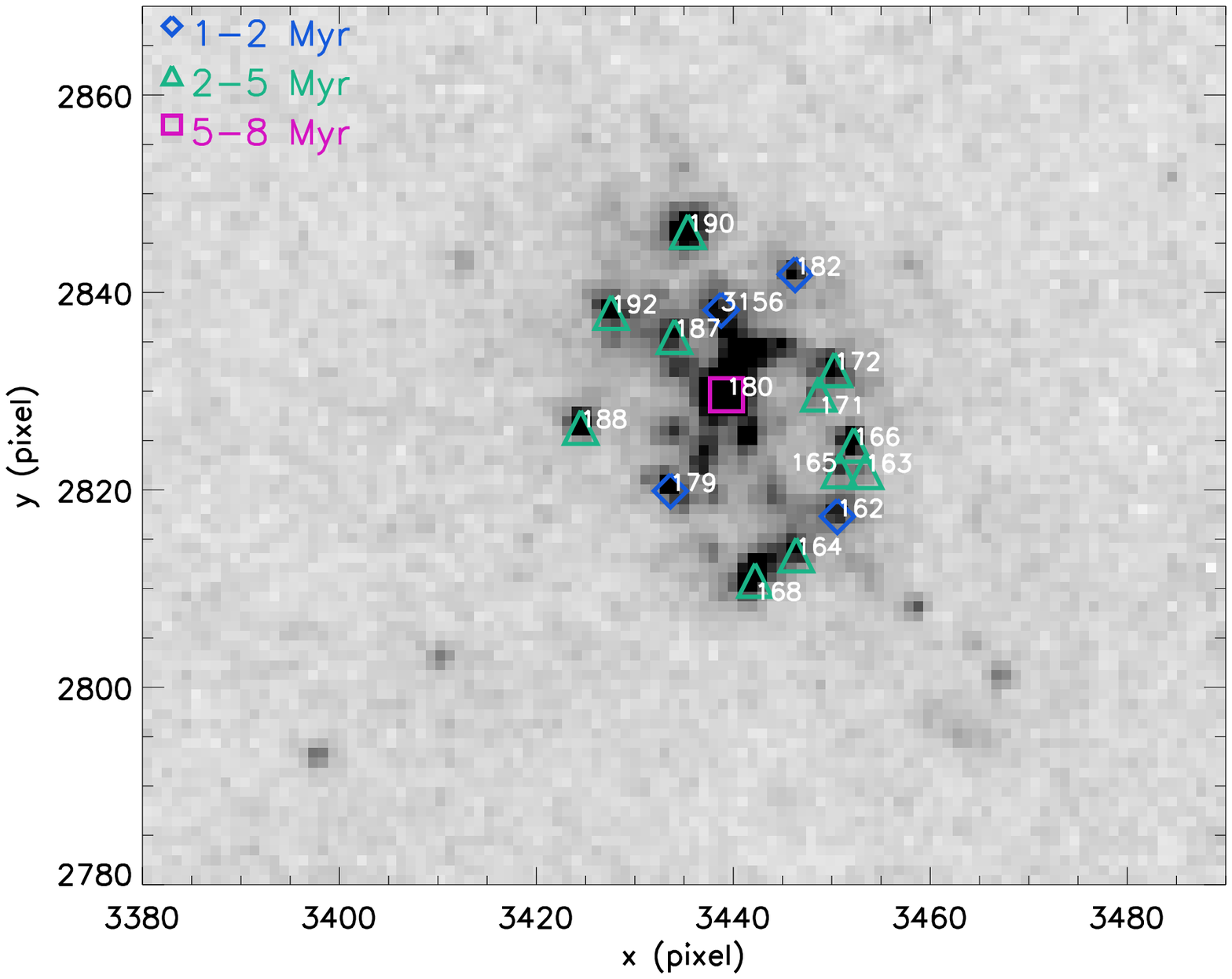}}}
\resizebox{0.49\hsize}{!}{\rotatebox{0}{\includegraphics{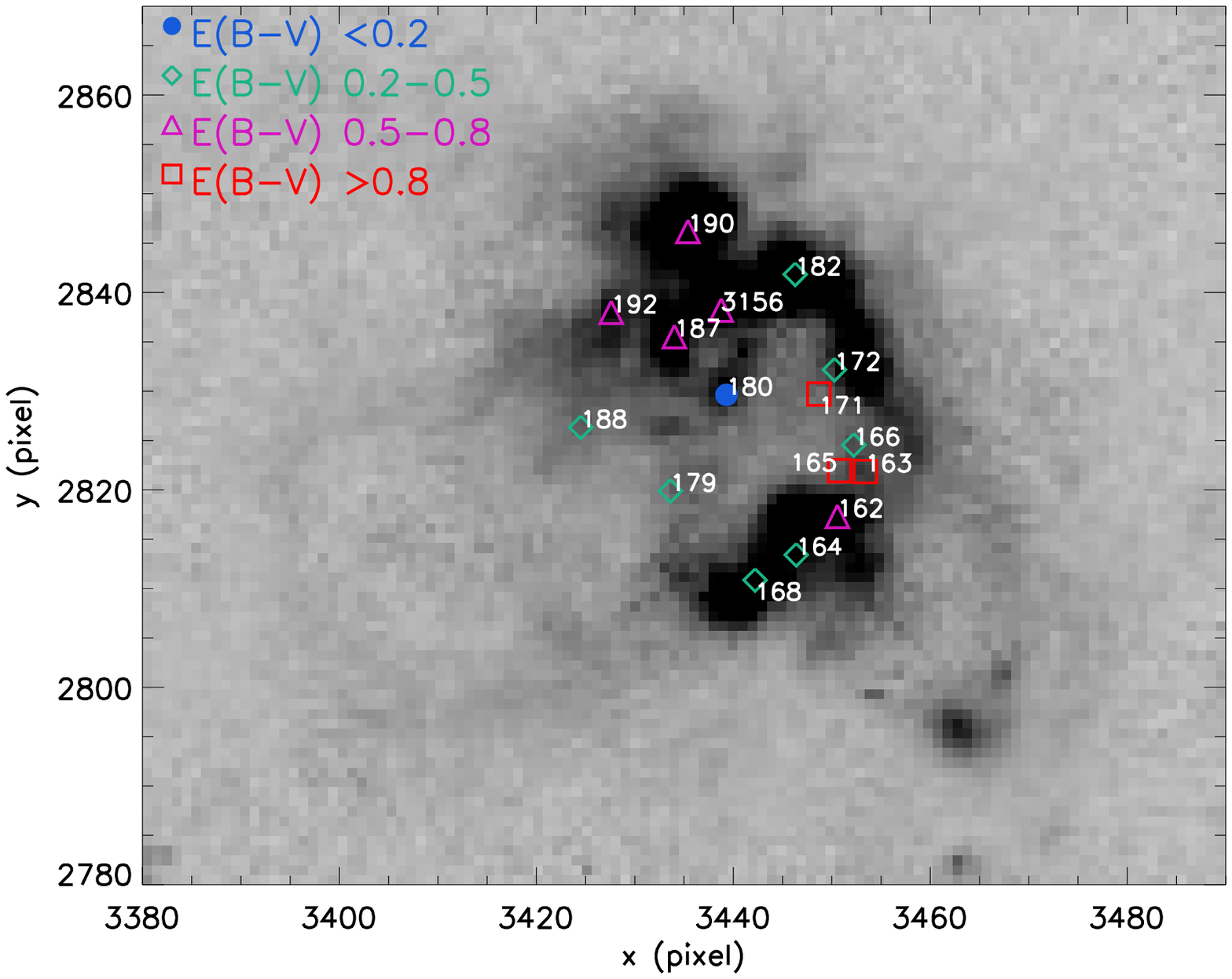}}}
\vspace{-10mm}
\caption{Left: the cluster ages in color code are overplotted to the F336W image of the ring. Right: the cluster visual extinction is overlaid on the F658N image, instead. The size of the 2 images is of 2.5" by 2.4" ($\sim130\times125$~pc$^2$). Size and orientation are the same as the inset in Figure~\ref{3col}.}
\label{age}
\end{figure*}

From Table~\ref{table1}, we see that the cluster masses are in the range $10^{3}$ and $10^4 \msun$, placing them into the stochastic regime \citep[e.g.][]{2004A&A...413..145C, 2009AJ....138.1724P, 2010A&A...521A..22F}. \citet{2011A&A...529A..25S} analysed the effect of stochastic IMF sampling in the recovered colors and ages of synthetic cluster populations of increasing stellar masses. In their Figure 8, for the mass range observed in the Ruby Ring complex, one can see that ages younger than 10 Myr are recovered for a large fraction of clusters in the age range 10-200 Myr, i.e. corresponding to the red supergiant phase.  However, since all the clusters in the ring have associated H$\alpha$ in emission, we are confident of their recovered ages. To estimate the uncertainties on the mass, we looked at the prediction of ${\cal M}^{min}$,  the mass value below which the observed cluster colors are severely biased with respect to the predictions of spectral evolutionary models \citep{2004A&A...413..145C}. The expected ${\cal M}^{min}$ in the $U$-band (our reference frame) is $\sim 3 \times 10^3 \msun$ at ages younger than 10 Myr, fairly close to our estimates. On the other hand, there are new indications (from globular cluster size and, independently, from the detection of the red giant tip, Adamo et al. in prep.) that the distance of the galaxy is higher than 11 Mpc, i.e. $\sim 18$ Mpc. If the galaxy is approximately at 18 Mpc, then, the correct cluster masses are almost a factor of 3 larger than the ones shown in Table~\ref{table1},
making them less affected by stochastic effects. For consistency, we use the distance of 11 Mpc and assume that the total stellar mass in clusters and all the derived estimates are upper limits to the real values. If the real distance is higher, our conclusions will not be significantly affected.

We find that the clusters of the Ruby Ring complex have ages younger than 10 Myr. In Figure~\ref{age}, left panel, the derived age ranges of the clusters are overplotted on the F336W image of the Ruby Ring, with different colors. In the right panel, we show the recovered extinctions, E(B-V), overplotted on the F658N image. From Fig. 2, the central cluster appears to be both older and less extinguished than the clusters in the ring.

The morphology of the ring changes quite drastically from the ultraviolet to H$\alpha$. The $U$-band reveals the sources of the ionising flux, i.e. numerous very young clusters with masses above $10^3 \msun$. The F658N transmits mostly the emission from the photoionized gas surrounding the clusters, highlighting the spatially extended H{\sc ii}-regions. An apparent outflow or "blow-out" region is seen to the lower left in Figure~\ref{age}. The cluster extinction appears to be lower in the region of the ring where we see evidence for an outflow in H$\alpha$. On the right edge and upper part of the ring, the extinction increases to values that are overall in agreement with the clusters being young and most likely partially embedded \citep[e.g.][]{2010MNRAS.407..870A}. 

\begin{figure}
\resizebox{\hsize}{!}{\rotatebox{0}{\includegraphics{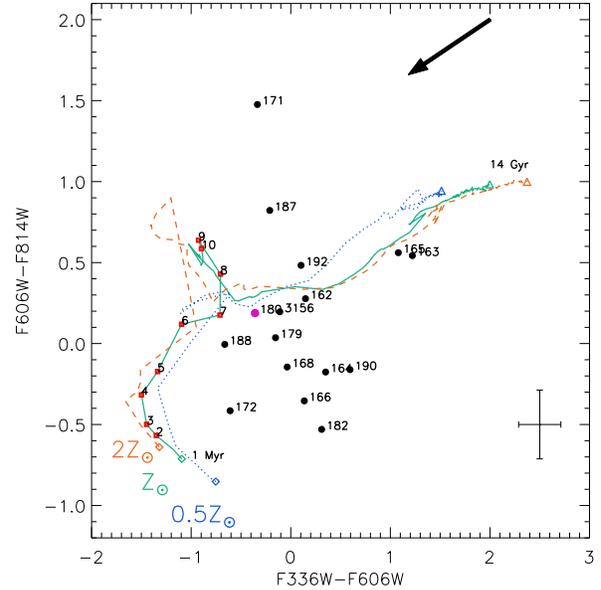}}}
\vspace{-6mm}
\caption{The F336W-F606W versus F606W-F814W color diagram of the Ruby Ring cluster complex. The clusters are shown as dots.  Their positions have not been corrected for extinction. The arrow indicates the direction and intensity of a de-reddening of E(B-V) $\sim$ 0.3 mag.}The central cluster 180 is plotted in purple. Three different metallicity evolutionary tracks by Yggdrasil models \citep{2011ApJ...740...13Z} are included in the plot. The red dots on the solar Z track (green solid line) indicate the position on the track at increasing ages (up to 10 Myr). The arrow indicates the direction and intensity of a de-reddening of E(B-V) $\sim$ 0.3 mag. The error bar shows the mean photometric error associated with the clusters. Clusters 171 and 187 have errors in F606W-band photometry much larger than 0.3 mag (limit for the $\chi^2$ analysis, see text), therefore their position in the diagram is rather uncertain.
\label{col-diag}
\end{figure}

 In Figure~\ref{col-diag}, we show the position of the clusters not corrected for intrinsic extinction in a color-color diagram. An arrow indicates in which direction the position of the points changes if corrected for extinction. We see that, except for a few outliers (large uncertainties in the F606W band photometry for clusters 171 and 187), the position of the clusters is compatible with an age spread of 1 to 7 Myr and variable extinction.  The best age and extinction of the central cluster 180 is of  7.0$^{+86.0}_{-0.0}$ Myr and E(B-V)$=0.1^{+0.01}_{-0.1}$.  The derived high upper age uncertainty is driven by the proximity of the colors of this cluster to the model predictions at that age together with a modest detected H$\alpha$ flux, which suggests that the cluster has already cleared its surroundings of the natal cold gas (compatible with an age $\sim7$ Myr). However, it is known that the radius of the H{\sc ii}-region from which H$\alpha$ is emitted increases as the cluster ages due to stellar feedback. Since we use fixed apertures to do photometry, it is possible that we are missing the bulk of the H$\alpha$ flux coming from 180. If this is the case, assuming that the extinction in the central region is much higher than obtained from the SED fitting technique, we expect 180 to have an age comparable to the clusters in the ring. We exclude, however that 180 has an age between 8 and 50 Myr, since its colors are not compatible with the location of the loop on the evolutionary track (see Figure~\ref{col-diag}). The SED fitting analysis of 180 shown in the Appendix confirms this scenario. The contour enclosing the 68 \% confidence limit of the derived properties of cluster 180 are two narrow regions centred around 7 and 70 Myr.

\begin{figure}
\resizebox{\hsize}{!}{\rotatebox{0}{\includegraphics{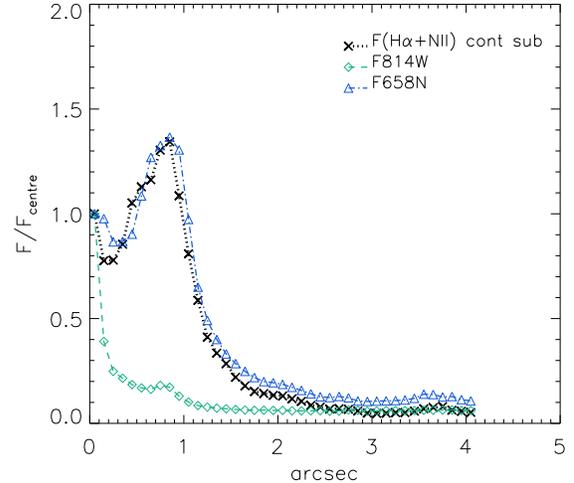}}}\\
\vspace{-6mm}
\caption{Surface density profiles of the Ruby Ring normalized to the flux of the core (coincident with the position of cluster  180). The black dotted curve has been obtained from the continuum subtracted F658N narrow filter and shows the profile of the H$\alpha + $ [N{\sc II}] flux density throughout the cluster complex. The other two lines (green dashed and blue dotted) show the profiles in the F814W and F658N filters, from which the continuum subtracted H$\alpha + $ [N{\sc II}] image has been obtained. The ring structure is clearly observed at about 0.8-0.9 arcsec (between 43 and 48 pc at the adopted galaxy distance).}
\label{halpha}
\end{figure}

As a final test we examined the surface density profile of the  continuum-subtracted H$\alpha +$ [N{\sc II}] flux across the Ruby Ring complex (Figure~\ref{halpha}). A continuum subtracted H$\alpha$ image has been produced using the two ACS/WFC F658N and F814W frames. The continuum at the F814W band is similar enough to the continuum at the H$\alpha$ line to give a fair representation of the resulting continuum-subtracted emission F(H$\alpha + $ [N{\sc II}]). However, we stress that any quantity derived from these flux measurements has to be considered an upper limit to the real values. We used the continuum-subtracted F(H$\alpha + $ [N{\sc II}]) image to produce surface brightness profiles throughout the cluster complex. In Figure~\ref{halpha} we show profiles normalized to the flux of the corresponding central region, coincident with the position of cluster 180. The continuum-subtracted H$\alpha$ emission is directly proportional to the amount of ionizing flux produced by the massive short-lived stars in the clusters and, as such, it is correlated to the stellar mass and anti-correlated with the cluster age. Using the derived masses of the clusters we estimate that the stellar mass per square parsec in the central circle (radius $\sim$ 2 px $\sim$ 0.1 arcsec $\sim$ 5.2 pc) is $\sim 60 \msun$/pc$^2$. Similarly, if we approximate that the stellar mass in clusters of the ring is distributed in an annulus corresponding to the brightest values of the profile in Figure~\ref{halpha} (between 0.75 and 0.85 arsec $\sim$ 40--45 pc) we find a mass surface density of 70 $\msun$/pc$^2$. The stellar mass in clusters per square parsec are comparable in the central region and in the ring. If the stellar population in the central region would be as young as that in the ring we would expect to see similar values of the F(H$\alpha + $ [N{\sc II}]) profile in the two regions (black dotted line) and a decline outside the complex. The shape of the F(H$\alpha + $ [N{\sc II}]) profile suggests that a young source, 180,  has swept up its own H{\sc ii}-region creating a cavity in the central area of the complex (central peak followed by a decline). A second prominent peak, about 70 \% brighter than the central one, is observed at the projected distance of 0.8-0.9 arcsec from the centre, i.e. at the location of the ring. This trend supports the evidence obtained from the SED fitting analysis, i.e. there may be an age spread between the central and the ring clusters.

It could be argued that the H$\alpha$ emission observed in the ring can be powered by the central cluster. In the following section we will see that the total H$\alpha$ luminosity, L(H$\alpha$), is comparable to starburst regions such as 30 Doradus in the Large Magellanic Cloud.  The amount of ionizing Lyman photons, $Q_0$, necessary to produce the H$\alpha$ luminosity observed in the Ruby Ring  requires a young stellar population ($<$ 4 Myr) with a mass of about $0.5-1\times10^5 \msun$. Massive stars are the main source of Lyman photons, therefore, the given age limit of 4 Myr is related to their short life-time. The central cluster in the Ruby Ring is too small (a factor of 10 in mass) and too old to be the main ionizing source in the ring. On the other hand, in the next section we will show that the total stellar mass derived for the clusters in the ring is compatible with being the main source of the observed L(H$\alpha$).

\subsection{Star formation rate and cluster formation efficiency in the Ruby Ring}
\label{ciaociao}

Using the continuum subtracted H$\alpha$ image, we derived the total luminosity, L(H$\alpha$), in the region of the Ruby Ring applying the following procedure. To correct for the contribution of the [N{\sc II}] lines, we used the predicted [N{\sc II}]  line ratio for case B recombination \citep{2006agna.book.....O} and the ratio of NGC\,2146 observed fluxes of H$\alpha$ and [N{\sc II}]$_{\lambda,6584.0}$ \AA, lines \citep{2006ApJS..164...81M}. We estimated the total F(H$\alpha$) inside a radius of 3 arcsec from the centre of the Ruby Ring, removing the F([N{\sc II}]) as a function of the transmission efficiency of the F658N filter at the position of the lines and multiplying by the width of the filter (73 \AA). L(H$\alpha$) was obtained by correcting for the foreground extinction  \citep{1998ApJ...500..525S} and internal extinction (average cluster extinction). We found L(H$\alpha$)$=4.6\times 10^{39}$ erg/s and derived \citep[applying the relation from][]{1998ARA&A..36..189K} a local star formation rate (SFR) and SFR surface density for the Ruby Ring of SFR $=0.037 \msun$/yr and  $\Sigma_{\textnormal{SFR}}=0.47 \msun$/yr/kpc$^{2}$, respectively. These values are quite similar to the ones found for cluster complexes in other nearby star-forming galaxies like M\,51 and the Antennae \citep{2005A&A...443...79B, 2006A&A...445..471B} and categorise the Ruby Ring as a localised starburst region. The total H$\alpha$ luminosity of the Ruby Ring is $\sim 1/3$ of the L(H$\alpha$) observed in 30 Doradus \citep{1984ApJ...287..116K}. 

Using the number of ionizing photons in the Lyman continuum per second emitted by a O7{\sc v} star \citep{2005A&A...436.1049M}, we find that the observed Ruby Ring  L(H$\alpha$) is consistent with being produced by $\sim 600$ O7{\sc v} stars. Summing up the derived (from the SED fitting analysis) stellar mass of each cluster in the complex,  the total stellar mass in clusters of the Ruby Ring is of $\sim10^5 \msun$. Integrating over a Salpeter IMF, the expected number of stars more massive of an O7{\sc v} star ($\geq 25 \msun$) is $\sim 200$. The agreement is within a factor of 3. However, we have to take into account that the  observed L(H$\alpha$) is an upper limit to the real value (see Section~\ref{age_sect}). Therefore, we conclude that the two measurements are comparable within the uncertainties.

From the total stellar mass of the star clusters in the Ruby Ring, assuming that it has formed in a time interval of 10 Myr (according to the age interval of the star clusters),  we obtain a formation rate of stars that form in clusters (CFR) of 9.6$\times 10^{-3}$ $\msun$/yr. Following \citet{2008MNRAS.390..759B}, who defined the cluster formation efficiency ($\Gamma=$CFR/SFR) as the fraction of star formation happening in clusters, we derived for the Ruby Ring $\Gamma = 0.38$.  The cluster mass has been derived assuming a Kroupa's IMF. Therefore, a factor has been applied to correct the total stellar mass  in clusters for the missing fraction if a Salpeter IMF would have been used, instead. This step is necessary because the calibration of the \citet[][]{1998ARA&A..36..189K} relation has been derived assuming a Salpeter IMF. The cluster formation efficiency is very high, comparable to starburst galaxies \citep[e.g.][]{2011MNRAS.417.1904A}, but we have also to take into account that we are looking at a very small region with numerous clusters, therefore biased toward the highly efficient star formation scales.

\section{The formation of the Ruby Ring}

The "collect and collapse" scenario predicts that an expanding H{\sc ii} region sweeps up the surrounding remaining low density material in a shell. The latter will eventually fragment and collapse, forming new stars. This scenario does not exclude that the shell may form in pre-existing molecular gas surrounding the H{\sc ii} region \citep{1998ASPC..148..150E}.

Recent simulations seem to reduce the role of the ionising front in triggering star formation in shells. \citet{2011MNRAS.414..321D} observe that the ionizing flux, produced by the massive stars formed in the densest gas regions, does not affect the cold gas regions, which will, eventually, become unstable and form a new generation of stars. A dense ionized gas  fills the voids created by the collapsing filamentary structure,  shielding the remaining surrounding filamentary structures, which will reach independently the critical conditions for star formation. \citet{2011arXiv1109.3478W}, observe, instead, that the location of dense star-forming clumps in the shells corresponds to the position of pre-existing cloud structures and that it does not require any collect and collapse mechanism to be formed.

As presented in Section~\ref{age_sect}, the age of the central cluster 180 is $\sim$7 Myr, while the mean age of the cluster ring is $\sim 3$ Myr. From the cluster analysis, we found that only 5 \% of the total derived stellar mass in clusters is contained in 180. The time delay between the formation of the central cluster and the ring structure, their relative mass ratio, and the regular morphology are all observables predicted by the "collect and collapse" triggering mechanism.  On the other hand these properties are challenging to explain with other scenarios which contemplates an uncorrelated formation of the central cluster and of a structured ring.  If feedback would have not  shaped the surrounding gas and star formation would have taken place in high gas density regions, we would have expected to observe a fractal morphology, typical of the collapsing GMCs. The presence of a central source, a cavity, and an well-shaped region where star formation has taken place after a lapse of time suggest a triggering event. Therefore, we will use the "collect and collapse" scenario to try to understand how the Ruby Ring has formed. 

\begin{figure}
\resizebox{0.8\hsize}{!}{\rotatebox{0}{\includegraphics{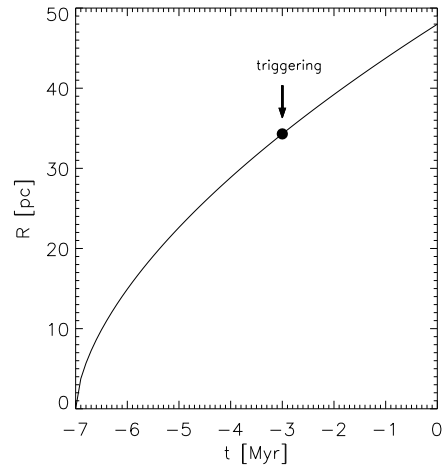}}}\\
\vspace{-6mm}
\caption{A representation of the shell expansion (radius, R) which formed the cluster ring, versus the time before the present. The spherical model by \citet{1998MNRAS.299..643E} has been used. The arrow indicates the radius and the time at which cluster formation was initiated in the ring. }
\label{shell}
\end{figure}

At the assumed distance of NGC\,2146, the radius of the ring, as obtained by Figure~\ref{halpha} is $\sim$ 48 pc (projected on the sky). If the ring has formed from the expanding bubble created by the central 180 cluster, then the current ring location should approximately spatially coincide with the position of the bubble. The latter has been created by continuous injection of energy from stellar winds and supernovae explosions produced in 180. We can derive the radius of the expanding shell using the equation for an adiabatic, pressure-driven bubble 
\citep[equation (21) in][]{1977ApJ...218..377W}. The equation depends on the mechanical luminosity produced by 180 and its age, and the mean density of the surrounding medium. To derive the mechanical luminosity, $L_{mech}$, we ran Starburst99 model \citep{1999ApJS..123....3L} for an instantaneous burst with an initial mass of  $5\times 10^3$ M$_\odot$, and determined the average value over the first 7 Myr. Assuming a typical Galactic number density (100 cm$^{-3}$) of the medium and 10\% efficiency for the conversion of the mechanical energy, we obtain a radius of $\sim$ 47 pc, which is in good agreement with the observed values.

According to the triggered scenario, however, the current position of the ring is not the same as the position at which the triggering event occurred. The second cluster generation has formed by swept-up expanding material. Due to the regular shape of the ring, we use the spherical solution for the analytical model of an expanding shell applied by  \citet{1998MNRAS.299..643E} to the Sextant region in the Large Magellanic Clouds. Using equation (26) and (28) by \citet{1998MNRAS.299..643E}, it is found that the ring formed at a smaller distance from the centre than the current location (see Figure~\ref{shell}). The velocity of the shell at the time the ring was formed is proportional to 3/5 of the ratio between the triggering radius ($\sim$34 pc) and the corresponding time (4 Myr), i.e. 5.1 km/s. We can verify that this derived expansion velocity (at the time the clusters in the ring were triggered) is in agreement with the expansion velocity driven by the mechanical energy injected by  180 when it was 3 Myr old. The latter can be estimated assuming a "snowplow expansion" (momentum and energy are conserved) of the wind bubble \citep{1999isw..book.....L} and using the $L_{mech}$ at the corresponding age ($\log L_{mech}$(3 Myr) $\sim38.13$ erg/s) and an ambient medium number density (100 cm$^{-3}$) as derived above. Applying equations 12.12 and 12.19 by \citet{1999isw..book.....L}, we obtain an expanding velocity of $\sim 3$ km/s, within a factor of 2 in agreement from the former estimate. 

The observed radius of the ring and the expansion velocity, derived from the properties of cluster 180, are in good agreement with the scenario where the formation of the ring has been triggered. Since the stellar mass traced in the ring structure is about a factor of 10 larger than in the central cluster, it is likely that the expanding shell encountered some pre-existing molecular gas. Therefore, the expanding ionising front passing through the cold gas cloud and the reverse shock formed after it, created the gravitational instability  necessary to trigger star formation in the ring. 

The environment where the Ruby Ring is located has most likely helped in shaping this extraordinary system. The Ruby Ring has formed at the end of one of the tidal streams. Star formation in the Ruby Ring's tidal stream started $\sim30-40$~Myr ago at its head and has continued until the present (Adamo et al. in prep.). We observe a spatial and temporal correlation in the clusters of the streams (older at the head and very young at the tail). It is not clear yet whether this sequentiality is due to a gravitational instability travelling throughout the streamed gas and it has only recently reached the extreme end of the ejected tail. The Ruby Ring has, thus, formed in a pocket of ejected gas located in a rather low density environment. The shape of the ring suggests that no shear is acting in deforming the system. This is, probably, the most important difference between the Ruby Ring and many galactic and extragalactic systems where triggering has produced arcs or asymmetric bubbles. The absence of shear has allowed the bubble produced by the central cluster to interact undisturbed and shape its surroundings. The availability of gas has allowed the formation of a ring of clusters.

\section{Conclusions}

We studied the physical properties and discussed the formation of the Ruby Ring system, a ring-like cluster complex in the starburst galaxy NGC\,2146. The Ruby Ring, so named due to its morphological appearance, is located in one of the tidal streams which surround the galaxy. NGC\,2146 is part of the Snapshot Hubble $U$-band cluster survey ({\it SHUCS}). The WFC3/F336W data has added critical information to the available archival Hubble Space Telescope imaging set of NGC\,2146, allowing us to determine ages, masses, and extinctions of the clusters in the Ruby Ring. We find evidence of a spatial and temporal correlation between the central cluster  and the clusters in the ring. The latter are about 4 Myr younger than the central cluster. This result is supported by the H$\alpha$ emission which is strongly coincident with the ring, and weaker at the position of the central cluster. From the derived total H$\alpha$ luminosity of the system we constrain the SFR density ($\Sigma_{\textnormal{SFR}}=0.47 \msun$/yr/kpc$^{2}$) to be quite high, comparable to the extended cluster complexes observed in M\,51 and the Antennae. The central cluster contains only 5 \% of the total stellar mass in clusters determined in the Ruby Ring. 

In order to explain the formation of the Ruby Ring, we considered the "collect \& collapse" scenario \citep{1977ApJ...214..725E}. We tested the predictions made by this model deriving the bubble radius and expansion velocity produced by the feedback from the central cluster.  We found that the location of the ring is quite similar to the determined position of the expanding bubble produced by the central cluster. At the time the star formation was triggered in the ring, the expanding velocity of the shell produced by the "collect \& collapse" model agrees within a factor of 2 with the derived velocity of the central expanding bubble. The "collect \& collapse" model is able to explain, at the same time, the morphology, the spatial and temporal correlation observed between the central clusters and the clusters in the ring, and the stellar mass ratio between the central source and the triggered ones. 

We conclude that, most likely, the Ruby Ring's remote location, at the end of the tidal stream, has created an almost unperturbed environment where the bubble produced by the central cluster has interacted and shaped its surroundings and, owing to the availability of gas, has triggered the formation of a ring of clusters.

\section*{Acknowledgments}
We thank the anonymous referee for suggestions and discussion which have improved the manuscript. Support for US participants in Program number SNAP-12229 was provided by NASA through a grant from the Space Telescope Science Institute, which is operated by the Association of Universities for Research in Astronomy, Incorporated, under NASA contract
NAS5-26555. The research leading to these results has received funding from the European Community's Seventh Framework Programme (/FP7/2007-2013/) under grant agreement No 229517. EZ acknowledges research grants from the Swedish Research Council and the Swedish National Space Board. This research has made use of the NASA/IPAC Extragalactic Database (NED) which is operated by the Jet Propulsion Laboratory, California Institute of Technology, under contract with the National Aeronautics and Space Administration.

\appendix
\section{SED fitting analysis}

\vspace{1cm}
\begin{figure*}
\resizebox{0.49\hsize}{!}{\rotatebox{0}{\includegraphics{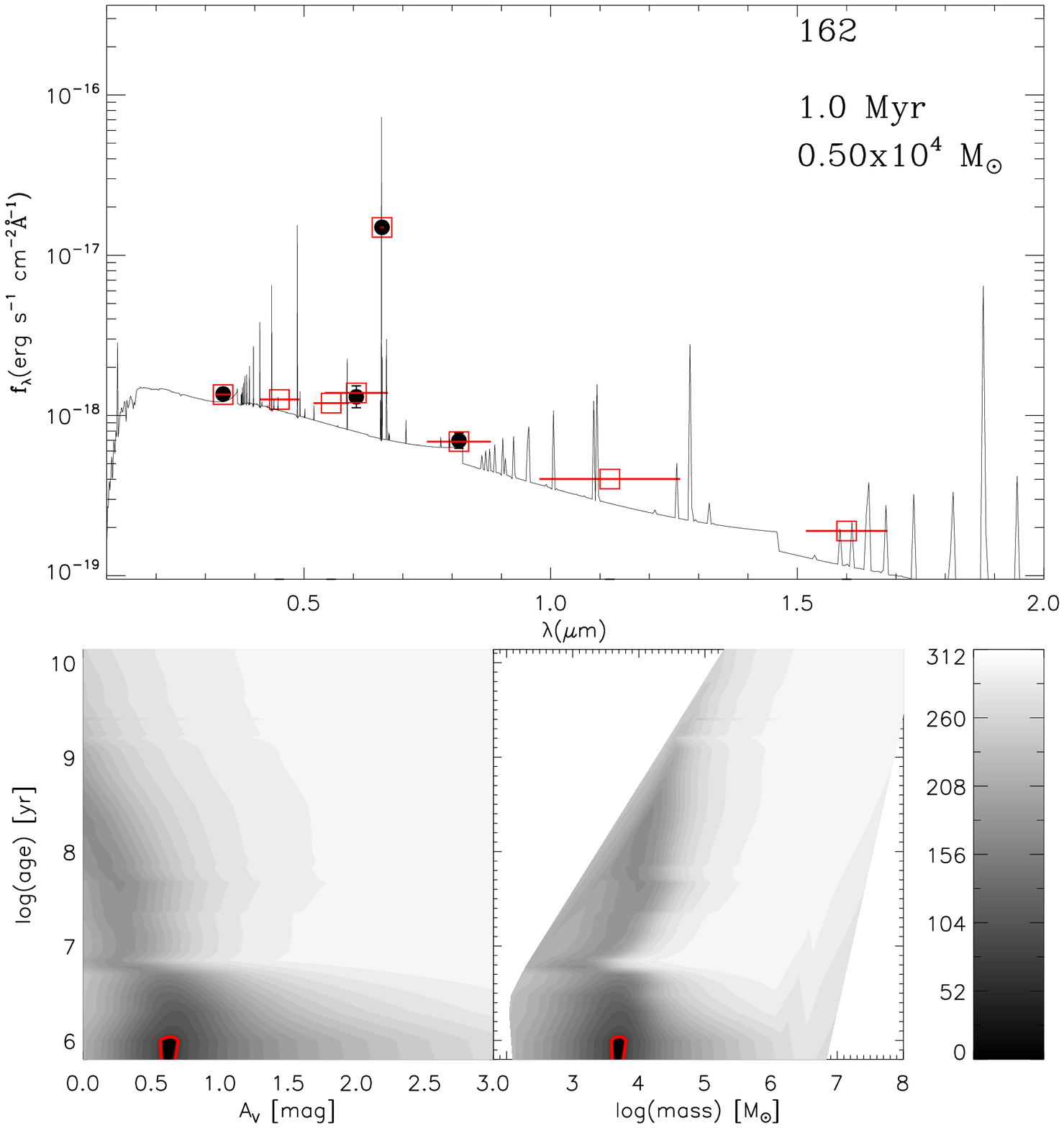}}}
\resizebox{0.49\hsize}{!}{\rotatebox{0}{\includegraphics{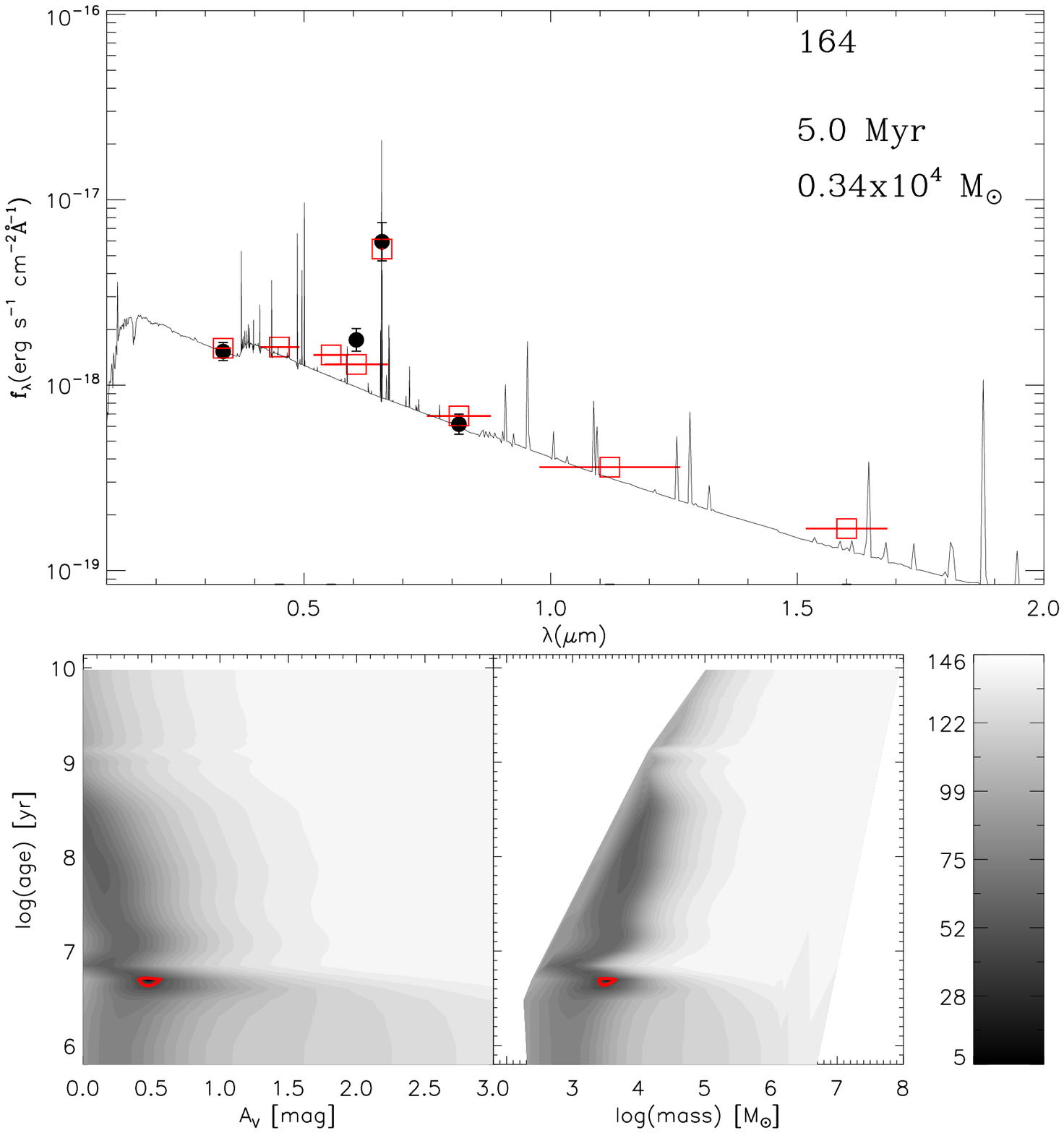}}}\\
\resizebox{0.49\hsize}{!}{\rotatebox{0}{\includegraphics{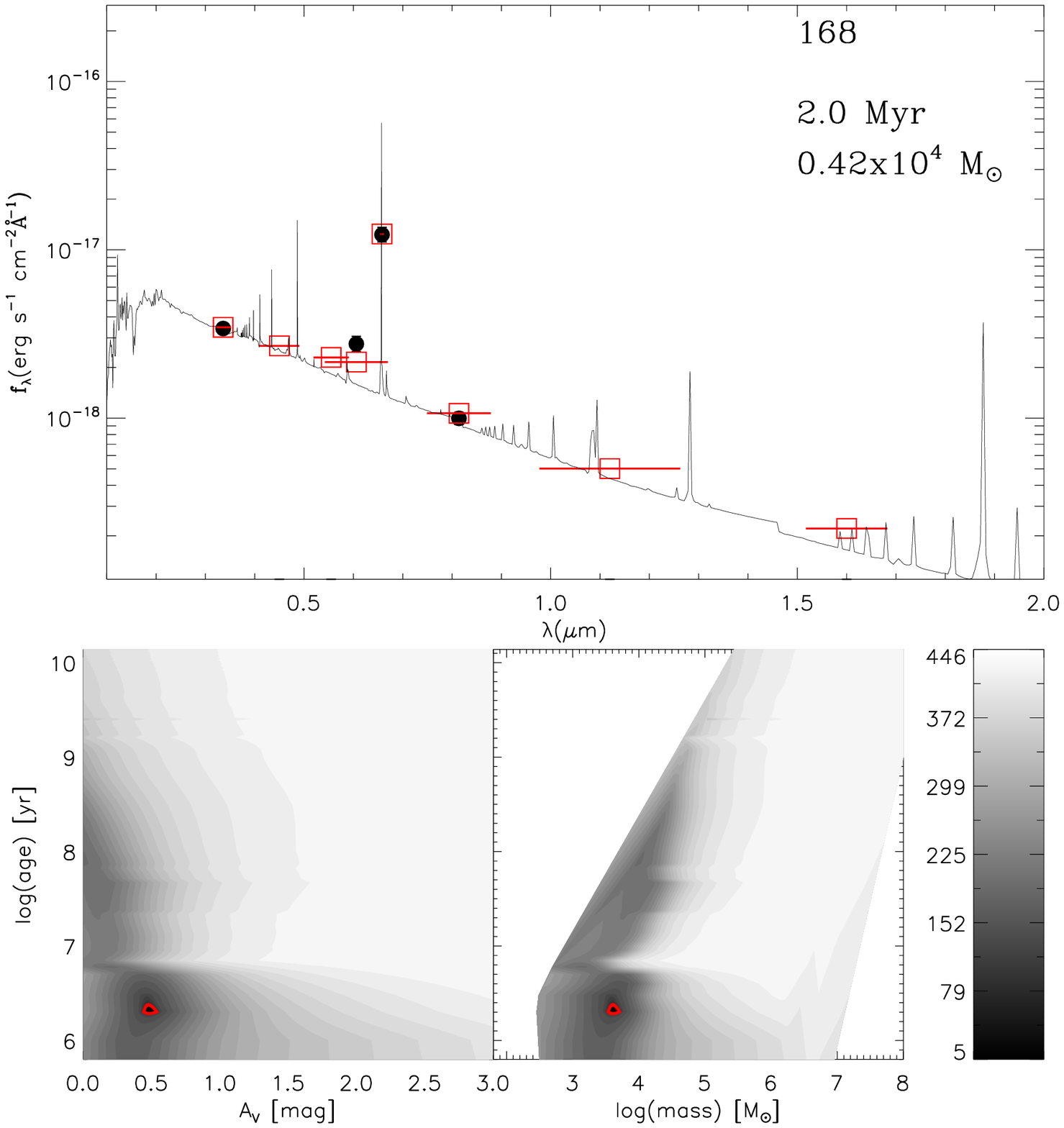}}}
\resizebox{0.49\hsize}{!}{\rotatebox{0}{\includegraphics{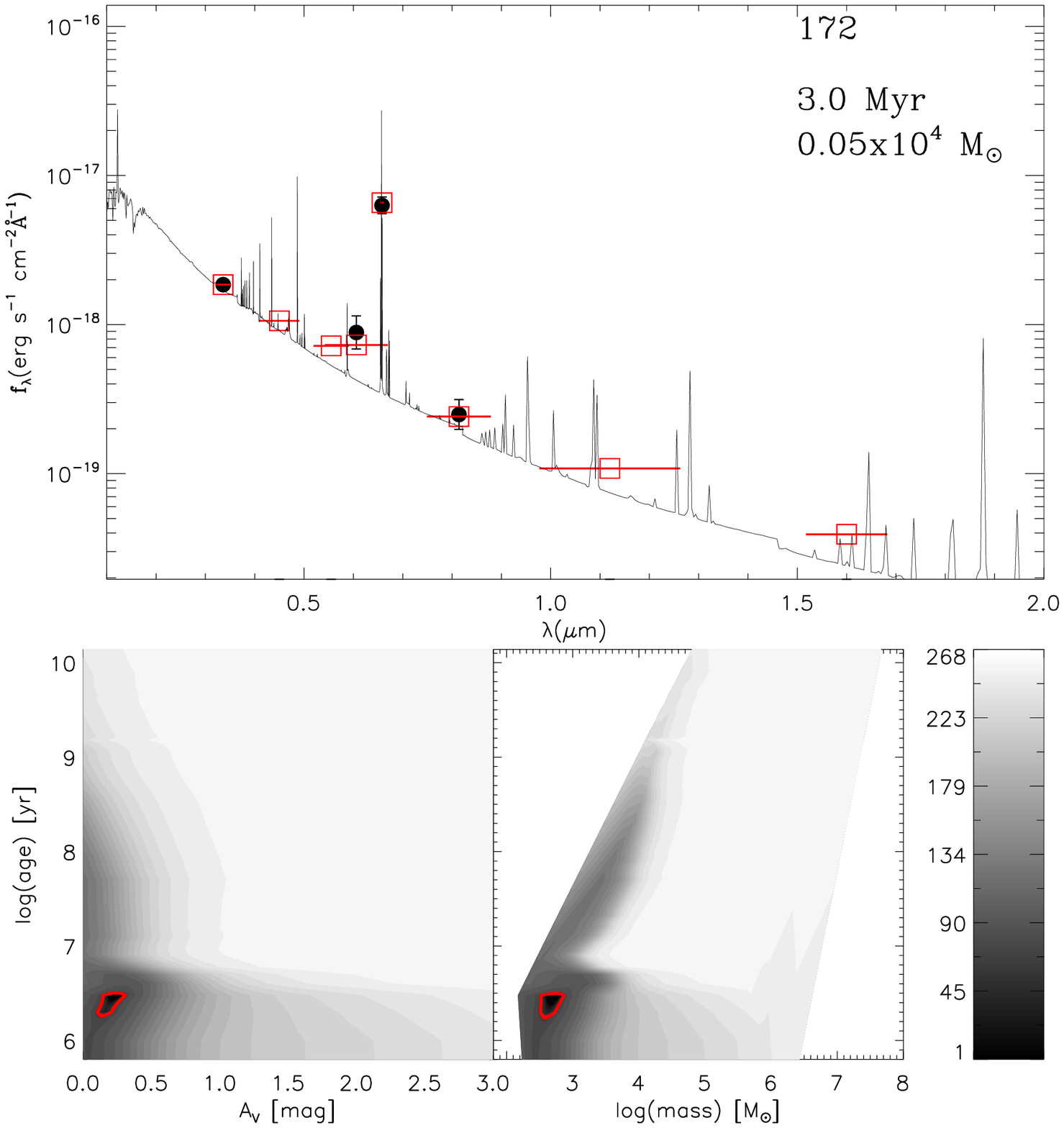}}}\\

\caption{The SED fitting analysis and error estimates for clusters in the Ruby Ring with detection in all the available filters. For each source, the top panel shows the quality of the SED fit. Black dots with error bars are detected fluxes, the red open squares are the integrated fluxes obtained convolving the underlying plotted model spectrum \citep[Yggdrasil models,][]{2011ApJ...740...13Z} with the corresponding filter throughput. The red horizontal bars indicate the width of each filter. The three bottom panels shows the parameter space and how the $\chi^2$ value changes as function of the input parameters. The bottom value of the bar corresponds to the best derived $\chi^2$. Lighter grey colors correspond to increasing large $\chi^2$ values. The red contour encloses the area corresponding to the 68 \% confidence limit.}

\end{figure*}

\begin{figure*}
\resizebox{0.49\hsize}{!}{\rotatebox{0}{\includegraphics{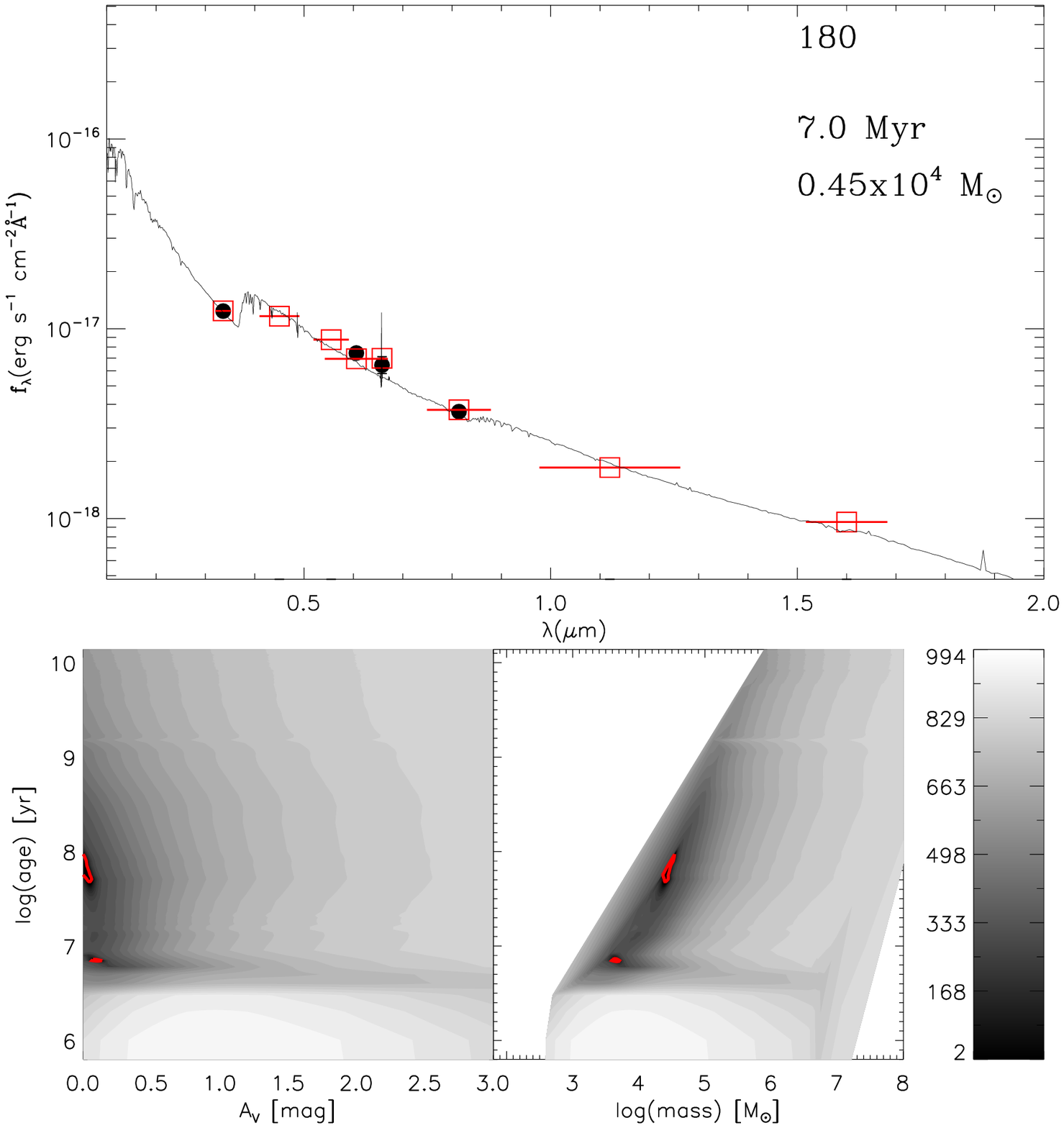}}}
\resizebox{0.49\hsize}{!}{\rotatebox{0}{\includegraphics{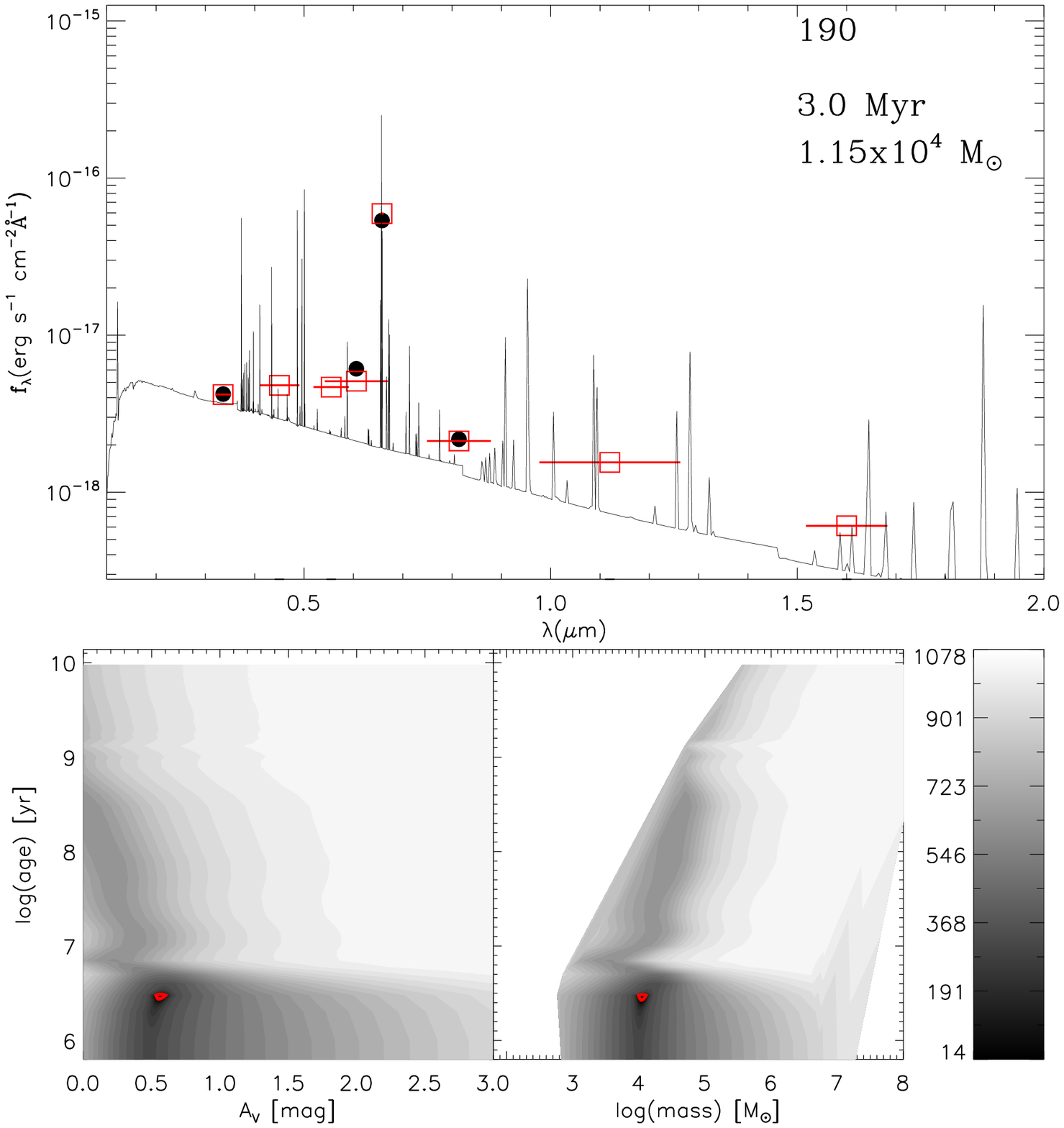}}}\\
\resizebox{0.49\hsize}{!}{\rotatebox{0}{\includegraphics{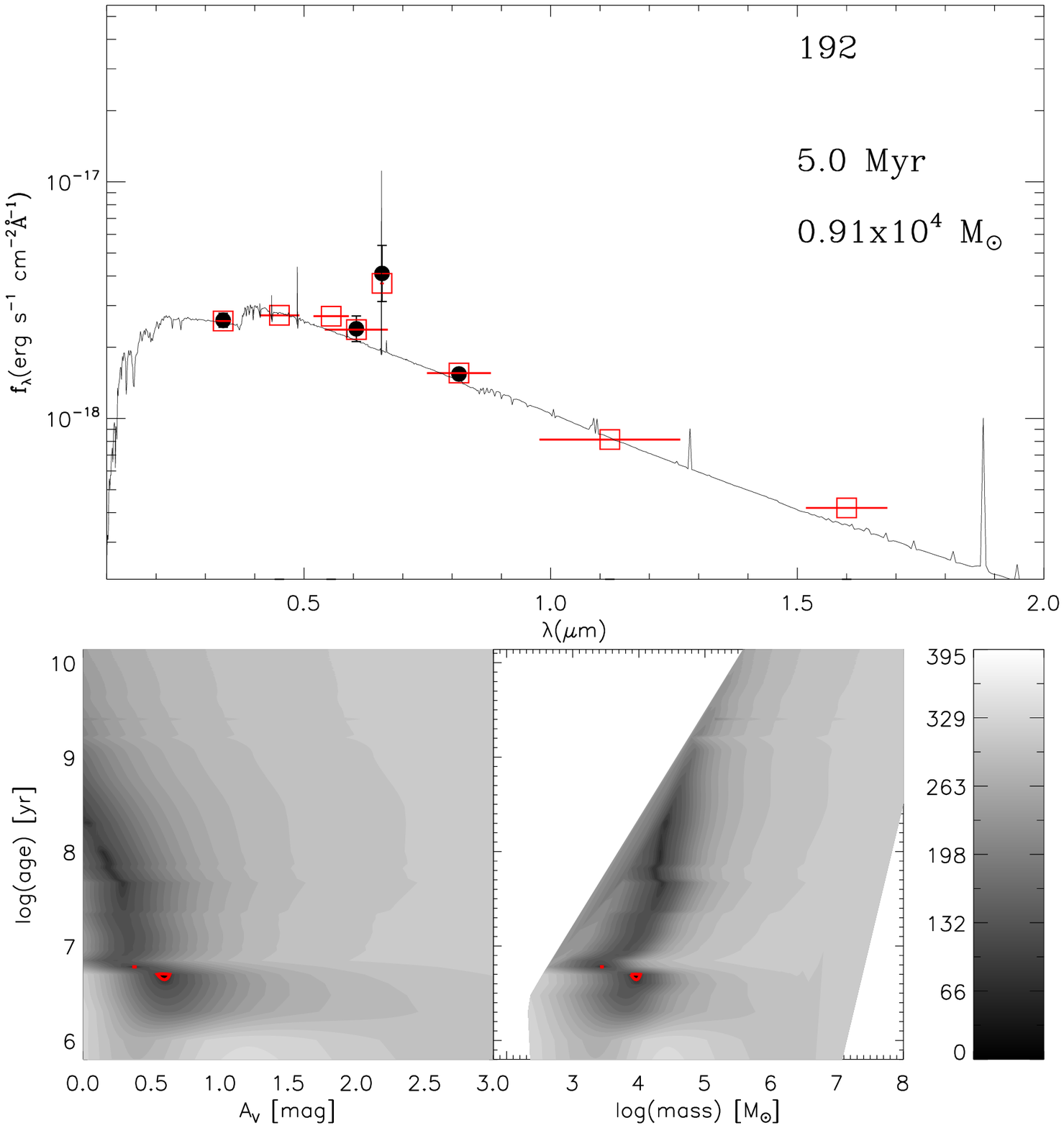}}}
\resizebox{0.49\hsize}{!}{\rotatebox{0}{\includegraphics{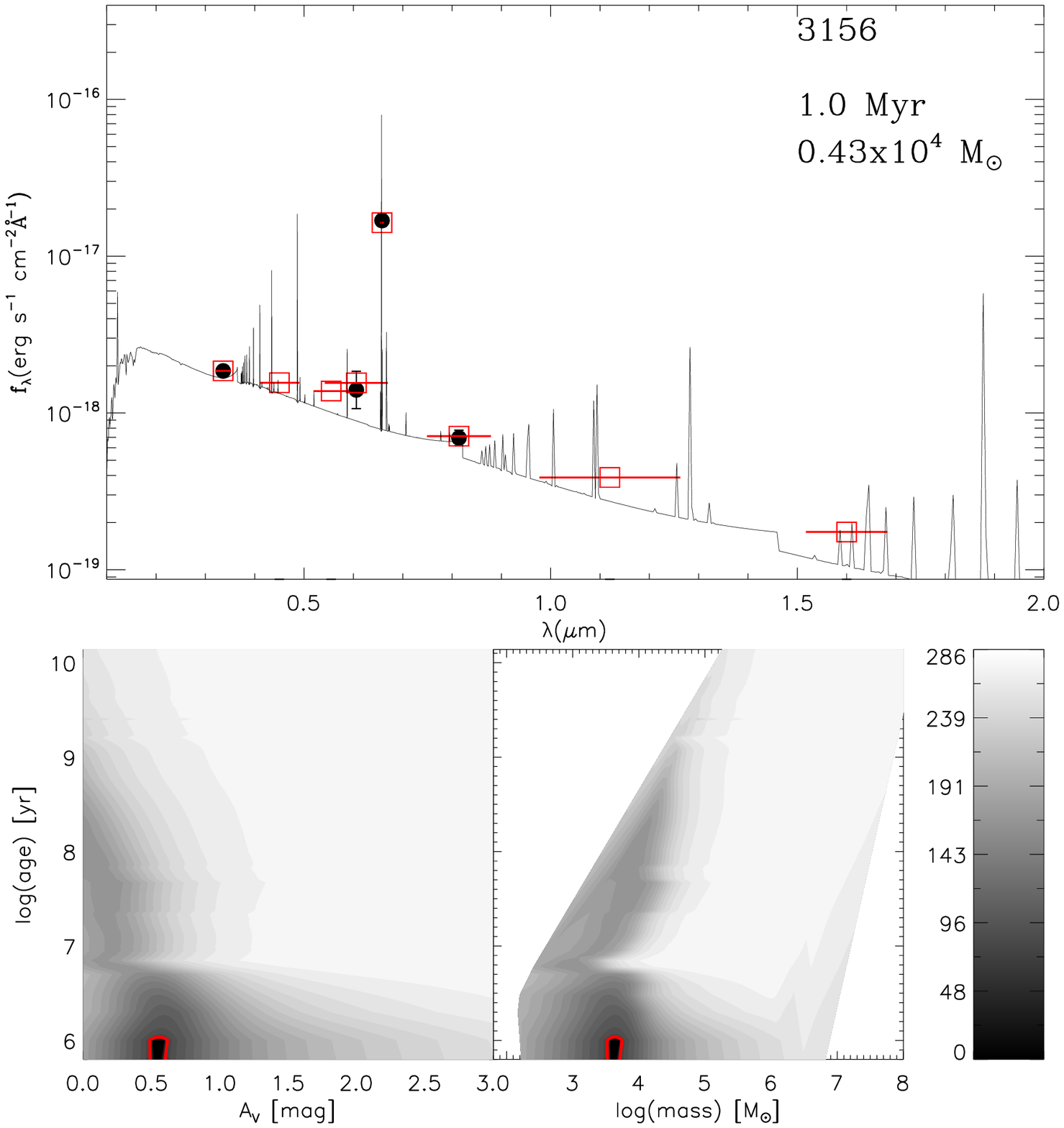}}}\\

\caption{The SED fitting analysis and error estimates for clusters in the Ruby Ring with detection in all the available filters. See previous figure's caption.}

\end{figure*}

\newpage

\newpage

\bsp

\label{lastpage}

\end{document}